\documentclass[%
 reprint,
 amsmath,amssymb,
pra,
floatfix,
]{revtex4-2}

\usepackage{graphicx}
\usepackage{dcolumn}
\usepackage{bm}
\usepackage{xcolor}
\usepackage{hyperref}
\usepackage{layouts}
\usepackage{multirow}

\begin{document}

\preprint{APS/123-QED}

\title{Resolving Power of Visible to Near-Infrared Hybrid $\beta$-Ta/NbTiN\\ Kinetic Inductance Detectors}

\author{Kevin Kouwenhoven$^{1,2}$}
\email{k.kouwenhoven@sron.nl}
\homepage{terahertz.tudelft.nl}
\author{Daniel Fan$^{3}$}
\author{Enrico Biancalani$^{5}$}
\author{Steven A.H. de Rooij$^{1,2}$}
\author{Tawab Karim$^{4}$}
\author{Carlas S. Smith$^{4}$}
\author{Vignesh Murugesan$^{1}$}
\author{David J. Thoen$^{2}$}
\author{Jochem J.A. Baselmans$^{1,2}$}
\author{Pieter J. de Visser$^{1}$}

\affiliation{%
 $^1$Netherlands Institute for Space Research (SRON), Niels Bohrweg 4, Leiden 2333 CA, Netherlands
 }%
\affiliation{%
 $^2$Department of Microelectronics, Delft University of Technology, Mekelweg 4 2628 CD, Delft, Netherlands
 }%
\affiliation{%
 $^3$Department of Precision and Microsystem Engineering, Delft University of Technology, Mekelweg 2, Delft 2628CD, Netherlands
 }%
\affiliation{%
 $^4$Delft Center for Systems and Control, Delft University of Technology, Mekelweg 2, Delft 2628CD, Netherlands
 }%
\affiliation{%
 $^5$Leiden Observatory, Niels Bohrweg 2, Leiden 2333 CA, Netherlands
 }%

\date{\today}

\begin{abstract}
Kinetic Inductance Detectors (KIDs) are superconducting energy-resolving detectors, sensitive to single photons from the near-infrared to ultraviolet. We study a hybrid KID design consisting of a beta phase tantalum ($\beta$-Ta) inductor and a NbTiN interdigitated capacitor (IDC). The devices show an average intrinsic quality factor $Q_i$ of 4.3$\times10^5$ $\pm$ 1.3 $\times10^5$. To increase the power captured by the light sensitive inductor, we 3D-print an array of 150$\times$150 $\mu$m resin micro lenses on the backside of the sapphire substrate. The shape deviation between design and printed lenses is smaller than 1 $\mu$m, and the alignment accuracy of this process is $\delta_x = +5.8 \pm 0.5$ $\mu$m and $\delta_y = +8.3 \pm 3.3$ $\mu$m. We measure a resolving power for 1545-402 nm that is limited to 4.9 by saturation in the KID's phase response. We can model the saturation in the phase response with the evolution of the number of quasiparticles generated by a photon event. An alternative coordinate system that has a linear response raises the resolving power to 5.9 at 402 nm. We verify the measured resolving power with a two-line measurement using a laser source and a monochromator. We discuss several improvements that can be made to the devices on a route towards KID arrays with high resolving powers.
\end{abstract}

\maketitle

\section{Introduction}

Kinetic Inductance Detectors (KIDs) \cite{day_broadband_2003} are superconducting detectors in which a single visible photon (1-3 eV) creates thousands of quasiparticle excitations through which we can measure the energy of each photon.
Compared to semiconductor-based detectors, where a visible photon excites just a few charge carriers, KIDs have zero dark current and read noise.
This makes them ideal candidates for photon-starved experiments such as exoplanet spectroscopy, which is one of the most appealing goals of astronomy in the coming decade.
Combined with their intrinsic colour resolution and photon arrival timing KIDs are promising detectors for chromatic wavefront sensors, multicolor fluorescence imaging, and order sorting detectors in a spectrometer.

A KID is a superconducting microwave resonator sensitive to a change in its complex conductivity, given by the kinetic inductance (Cooper pairs) and resistance (quasiparticles).
A signal with enough energy to break Cooper pairs, creating quasiparticles in the process, will change the resonator's inductance and resistance, shifting the resonance condition.
The energy gap of a superconductor is $<$ 1 meV, so an absorbed visible photon (1-3 eV) creates thousands of quasiparticles. 
The number of quasiparticles created, and thus the magnitude of the response, depends on the energy of the absorbed photon, making visible to near-infrared KIDs energy resolving detectors.

The exact number of quasiparticles created for a given photon energy is uncertain due to the statistical nature of the energy down-conversion process \cite{kozorezov_quasiparticle-phonon_2000, kozorezov_energy_2012}.
The statistical uncertainty of this process gives an upper limit to the attainable resolving power $(E/\delta E)$ \cite{kurakado_possibility_1982, rando_properties_1992} $R_{Fano} \approx (1/2.35)\sqrt{(\eta_{pb}^{max} E_{ph})/(\Delta F)}$ with $E_{ph}$ the photon energy, $\Delta$ the superconducting gap energy, $F=$ 0.2 the Fano limit for superconductors \cite{kurakado_possibility_1982,rando_properties_1992}, and $\eta_{pb}^{max}=$ 0.59 the pair breaking efficiency \cite{guruswamy_quasiparticle_2014,de_visser_non-equilibrium_2015,kozorezov_quasiparticle-phonon_2000}.

KIDs in Integral Field Units for exoplanet spectroscopy require a resolving power $R$ of $\sim$100.
To reach this resolving power, the ideal superconductor for a
 KID has a low critical temperature $T_c$ corresponding to a low gap $\Delta$.
Additionally, for absorber based KIDs \cite{meeker_darkness_2018,mazin_superconducting_2012,mazin_arcons_2013,szypryt_large-format_2017,zobrist_design_2019} where the incoming energy is absorbed directly in the inductor, a high resistivity ($\rho$) superconductor is essential for a high absorption efficiency \cite{kouwenhoven_model_2022}.
A higher resistivity superconductor will, for any KID, reduce the pixel size and increase the responsivity due to the higher kinetic inductance.

In the ongoing search for the ideal KID material \cite{mazin2020superconducting}, several materials have been reported including sub-stoichiometric TiN ($T_c$ = 0.8 K, $\rho$ = 100 $\mu\Omega$cm) \cite{mazin_superconducting_2012, mazin_arcons_2013}, PtSi ($T_c$ = 0.9 K, $\rho$ = 35 $\mu\Omega$cm) \cite{szypryt_large-format_2017} and Hf ($T_c$ = 0.4 K, $\rho$ = 97 $\mu\Omega$cm) \cite{zobrist_design_2019}.
One of the low $T_c$ elemental superconductors that remains unexplored is beta-phase tantalum ($\beta$-Ta) \cite{mazin2020superconducting} which has a critical temperature of around 0.6 - 1.0 K and resistivity of $\approx$ 200 $\mu\Omega$cm.
The other phase, $\alpha$-Ta \cite{face_nucleation_1987}, has a lower resistivity and a much higher $T_c$.

Here we demonstrate hybrid $\beta$-Ta/NbTiN KIDs, where radiation coupling to the beta-phase tantalum ($\beta$-Ta) inductor is done using a 3D printed microlens array printed directly on the chip backside.
The hybrid design consists of a light-sensitive, low-$T_c$ inductor and a high-$T_c$ capacitor.
The inductor is made out of $\beta$-Ta which has a $T_c$ of 1.0 K and a resistivity of 239 $\mu\Omega$cm.

The capacitor is made out of NbTiN with a $T_c$ of 15.2 K.
This design approach has been used in Al/NbTiN KIDs to confine the quasiparticles to the sensitive part of the detector \cite{janssen_high_2013, yates_photon_2011}.
Additionally, NbTiN is resistant to the Hf cleaning steps required for good galvanic contacts between the inductor and capacitor, and provides a practical readout circuit with galvanically connected bridges for high fidelity arrays. 

We show the resolving power of these devices from 402 to 1545 nm, limited by saturation in phase response from $R=$ 4.6 at 1545 nm to $R=$ 4.9 at 402 nm.
An alternative coordinate system that is linear in photon energy, presented in \cite{zobrist_improving_2021}, raises the resolving power at 402 nm to 5.9.
We demonstrate that the obtained resolving power agrees with a two-line resolving measurement.
Finally, we discuss the advantages, and current limitations, of our devices and present a route towards high resolving powers with $\beta$-Ta based KIDs. 



\section{Design and Fabrication}
\label{sec:Des_and_Fab}

\begin{figure}[h!]
  \includegraphics[width = \linewidth]{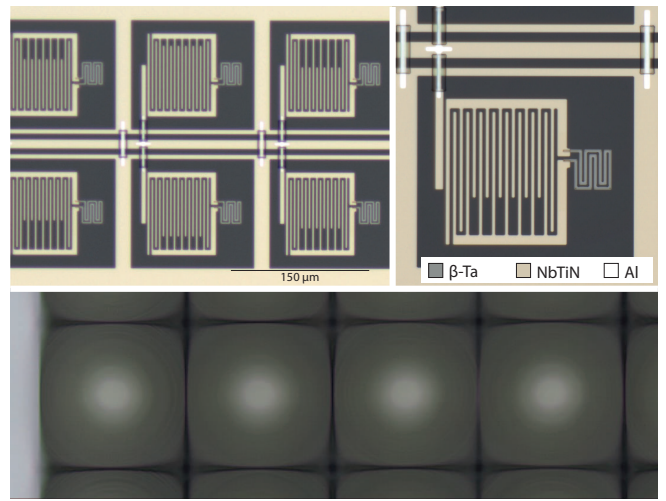}
  \caption{Microscope images of the NbTiN/$\beta$-Ta hybrid LEKID. \textbf{Left:} a small section of the 6$\times$6 array with the double side coupling bar geometry. The finger length of the IDC varies from KID to KID to achieve the desired frequency spacing. The pixel pitch is 150 $\mu$m. \textbf{Right:} The hybrid KID consists of a $\beta$-Ta meandered inductor and a NbTiN IDC on a sapphire substrate. \textbf{Bottom:} A section of the 3D printed 150$\times$150 $\mu$m lens-array. The lens array is printed on the backside of the 350 $\mu$m thick substrate. The lenses are designed with a height of 55 $\mu$m, and a conic constant of -0.405.}
  \label{fig:KID}
\end{figure}

We use a hybrid lumped element design (LEKID), consisting of a high resistivity $\beta$-Ta inductor and a NbTiN interdigitated capacitor (IDC).
The finger lengths of the IDC set the KID frequency spacing, see Fig.~\ref{fig:KID}.
Each KID capacitively couples to the NbTiN coplanar waveguide (CPW) readout-line through a NbTiN coupling bar, which runs alongside the IDC and galvanically connects to the central line of the CPW with an aluminium bridge on top of a polyimide dielectric support.
We use a double-sided coupling bridge, where one bridge connects to both KIDs at either side of the readout-line, see Fig.~\ref{fig:KID}.
A similar bridge design galvanically connects both CPW ground planes at regular intervals.
This design is similar to the one presented in \cite{meeker_darkness_2018}.
We have designed two chip variations, one containing 6$\times$6 hybrid KIDs as described here, and one containing 2$\times$6 hybrid KIDs with the other locations in the 6$\times$6 positions filled with design variations and lens alignment test structures.

We use a 350 $\mu$m thick c-plane sapphire substrate.
The IDC and readout-line are 150 nm of reactive magnetron sputter-deposited NbTiN \cite{thoen_superconducting_2017,bos_reactive_2017}.
The inductor is 60 nm of sputter-deposited $\beta$-Ta.
The deposition and etching order is NbTiN, polyimide, $\beta$-Ta, protection resist, and Al.
The resist cap protects the $\beta$-Ta inductor during the Al processing steps.
Removing the resist patch is the final step in the fabrication process.
Before each metal step, the wafer is cleaned with HF, which is critical for a good galvanic contact to the NbTiN layer.

We use a four-probe DC structure to measure the DC properties of each layer.
We measure a $T_c$ of 1.0 K, a normal resistance of $1595$ $\Omega$ just above the superconducting transition, and a room temperature (300 K) resistance of 1640 $\Omega$ which gives a residual resistivity ratio (RRR) value of 1.03.
This yields a sheet resistance $R_s$ of 39.8 $\Omega$ for the 60 nm thick $\beta$-Ta layer, resulting in a sheet kinetic inductance $L_k$ of 54.6 pH/$\square$.
For the ~120 nm NbTiN layer we measure $T_c$ of 15.2 K and $R_s$ of 11.0 $\Omega$.

The photon sensitive part of the LEKID is the inductor, which makes up $\sim$2\% of the total LEKID footprint.
We focus the light onto the inductor by printing an array of elliptical micro-lenses on the backside of the sapphire substrate.
The lenses are elliptical, instead of simply spherical, to reduce the effect of spherical abberations.
Each lens is designed to have a 150 $\mu$m square base, a height of 55 $\mu$m, and a conic constant of -0.405.
The lenses are printed with a Nanoscribe Photonic Professional GT2 Two-Photon Polymerization (TPP) printer in IP-S resin using a 25$\times$ immersion objective.
The printer aligns the lenses to markers etched in the NbTiN layer.
This method yields an alignment error of $\delta_x = +5.8 \pm 0.5$ $\mu$m and $\delta_y = +8.3 \pm 3.3$ $\mu$m, measured for 6 lenses.
The measured FWHM spot size at the detector level is 6.2 $\pm$ 0.7 $\mu$m at 673 nm.
The measurement procedures and a detailed analysis, including a profilometric measurement, are presented in Appendix \ref{supp:Lens_alignment}.

\section{Measurement setup}
\label{sec:Setup}

\begin{figure}[h!]
	\includegraphics[width = \linewidth]{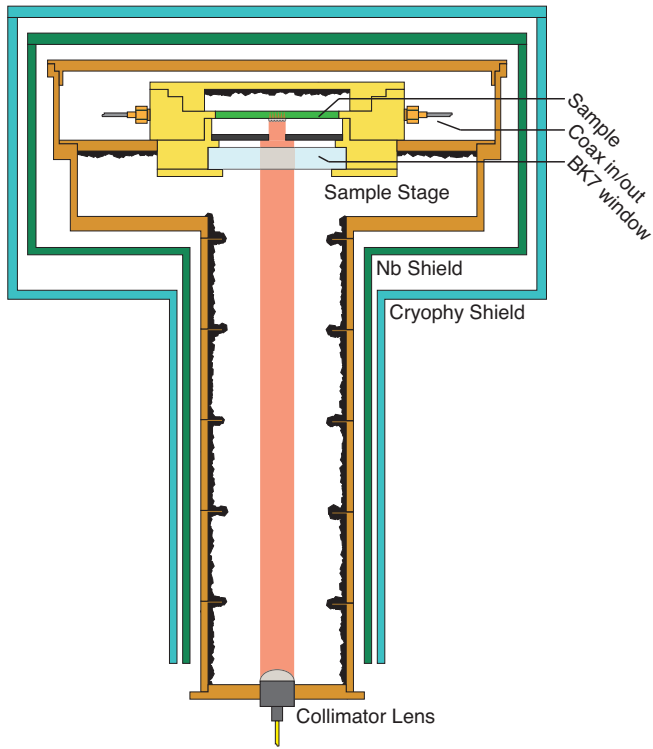}
  \caption{The 100 mK sample stage of the setup, parts are not to scale. Light enters the stage through a collimator lens connected to an optical fiber. The collimated bundle passes through a snout coated with black, carbon loaded epoxy. At the sample stage the bundle passes through a 5 mm thick BK7 window, and a 1 mm diameter aperture before reaching the sample. The 100 mK stage is surrounded by 2 magnetic shields, a superconducting Niobium shield, and a Cryophy shield.}
  \label{fig:Set-up}
\end{figure}

The samples are cooled down in a pulse-tube pre-cooled dilution refrigerator.
All measurements are performed at a base temperature of 100 mK.
We use a box-in-box sample stage design, conceptually similar to \cite{de_visser_fluctuations_2014,baselmans_ultra_2012}, to shield the sample from stray light coming from the 4K stage of the cooler.
Before reaching the sample, light travels through a corrugated snout and passes through a 5 mm thick BK7 glass window as illustrated in Fig.~\ref{fig:Set-up}.
Light from off-normal incidence is absorbed in the snout using black, carbon-loaded epoxy.

Located at the input of the snout is a collimator lens (Thorlabs F260FC-B) connected to an optical fiber (Thorlabs SMF-28-J9, with an 8.2 $\mu$m diameter core and 125 $\mu$m cladding, single-moded at 1550 nm).
The collimator has a 3 mm beam diameter at the sample location (633 nm), while the array size is constrained to a 1 mm diameter for uniform illumination of the array.
Alternatively, the lid with the collimator lens can be removed, providing optical access from outside the cryostat to the sample stage through a set of windows and filters.

The input of the fiber is either one of four fiber-coupled lasers (Thorlabs), a monochromator (Oriel Cornerstone 260) or the combination of the two through a fiber combiner (Thorlabs TW670R5F1).
The fiber-coupled lasers have wavelengths 1545, 986, 673, and 402 nm, with output power from 1-10 mW in continuous wave mode and a line-width corresponding to $R$ $>$ 300.
A combination of mechanical and digital attenuators set the laser power at the sample holder.

\section{Measurements and Analysis}
\label{sec:Meas}
\subsection{Resonator properties}
\label{subsec:Res_Prop}

To characterize the device, we perform a microwave measurement with a standard homodyne detection scheme, a detailed overview of the measurement and set-up can be found in \cite{de_visser_quasiparticle_2014}.
We perform a frequency sweep to obtain the resonance circle (S21) for each KID. Fitting the resonance dip $(|S_{21}|)$ to an analytical model \cite{de_visser_quasiparticle_2014,khalil_analysis_2012} provides the resonator's resonance frequency $f_0$ and quality factors.
We measure the resonance frequencies between 7.5 and 9.5 GHz, consistent with a SONNET simulation using the measured $L_k$ of 54.46 pH/$\square$ from the four-probe DC measurement.
For a chip with 12 hybrid $\beta$-Ta KIDs, spaced from 7.86 to 8.25 GHz, we measure an average $Q_i$ of 4.3$\times10^5$ $\pm$ 1.3 $\times10^5$. The average coupler quality factor $Q_c$ is 3.2$\times10^4$ $\pm$ 2.0$\times10^4$ for a designed $Q_c$ (at 4 GHz) of 2$\times10^4$.
The optimal read power of each KID is set manually by selecting a power just below the bifurcation point.
For powers at or above the bifurcation point, the resonator is operated in a non-linear regime where the resonance curve ($S_{21}$) shows hysteretic switching \cite{devisser_readout_2010,swenson_operation_2013}.

\begin{figure}[h!]
  \includegraphics[width = \linewidth]{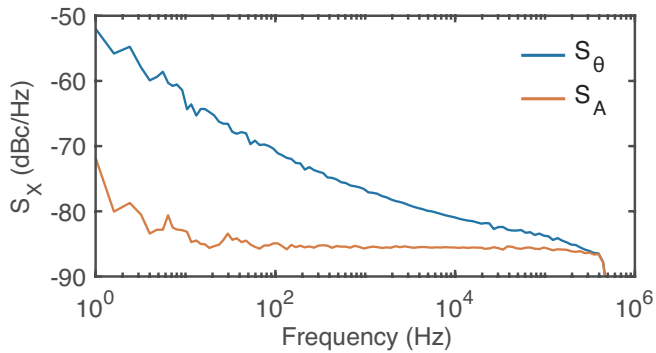}
  \caption{Measured Power Spectral Density (PSD) of phase and amplitude response for a hybrid $\beta$-Ta/NbTiN LEKID with $f_0 = 8.15$ GHz, $Q_i = 6.72\times10^5$, $Q_c = 1.88\times10^4$, KID internal power $P_{int} = -61$ dBm, and at a bath temperature of 100 mK. The measurement is performed without laser illumination.}
  \label{fig:S21_noise}
\end{figure}

We perform a noise measurement by taking two time-domain streams of the IQ-response (Appendix \ref{supp:coord}) at the resonance frequency, one of 40 seconds sampled at 50 ksample/s; and one of 0.5 seconds sampled at 1 Msample/s.
The IQ-response is translated to the polar KID coordinates: phase, and amplitude \cite{de_visser_quasiparticle_2014}. Fig.~\ref{fig:S21_noise} shows the resulting power spectral density (PSD).

\subsection{Resolving power}
\label{subsec:Res_Pow}

\subsubsection{Optimal filter}
We determine the energy resolution of a KID from the histogram of photon pulse heights, obtained from a laser measurement at a single wavelength.
For this measurement, we follow the approach described in \cite{irwin_phonon-mediated_1995,de_visser_phonon-trapping-enhanced_2021}.
We consider a photon pulse in the KID response $D(f)$ which can be modelled as 

\begin{equation}	
	\hat{D}(f)=H(E_{ph})M(f)+N(f),
	\label{eq:pulse_model} 
\end{equation}

where $f$ is the frequency and $D(f)$ the Fourier transform of the time domain signal $d(t)$.
The modelled photon pulse $\hat{D}(f)$ consists of a pulse shape $M(f)$ multiplied by a pulse height $H(E_{ph})$, which is a function of the photon energy $E_{ph}$, measured in the presence of noise $N(f)$.
The measured and model pulse can be defined in any of the KID readout coordinates \cite{de_visser_quasiparticle_2014}, we typically use the phase response $\theta$ unless specified otherwise: $D(f) = \theta(f)$. 
We assume that the pulse shape $M(f)$ and the noise $N(f)$ are not a function of the energy of the incoming photon $E_{ph}$.
The optimal filter for the pulse height $H(E_{ph})$ is then given by \cite{irwin_phonon-mediated_1995, eckart_measurements_2007}:

\begin{equation}
		\hat H(E_{ph}) = \int_{-\infty}^{\infty}\frac{D(f)M^*(f)}{|N(f)|^2}df \bigg/ \int_{-\infty}^{\infty}\frac{|M(f)|^2}{|N(f)|^2}df.
\label{eq:op_filter}
\end{equation}

\begin{figure*}[ht]
	\includegraphics[width = \textwidth]{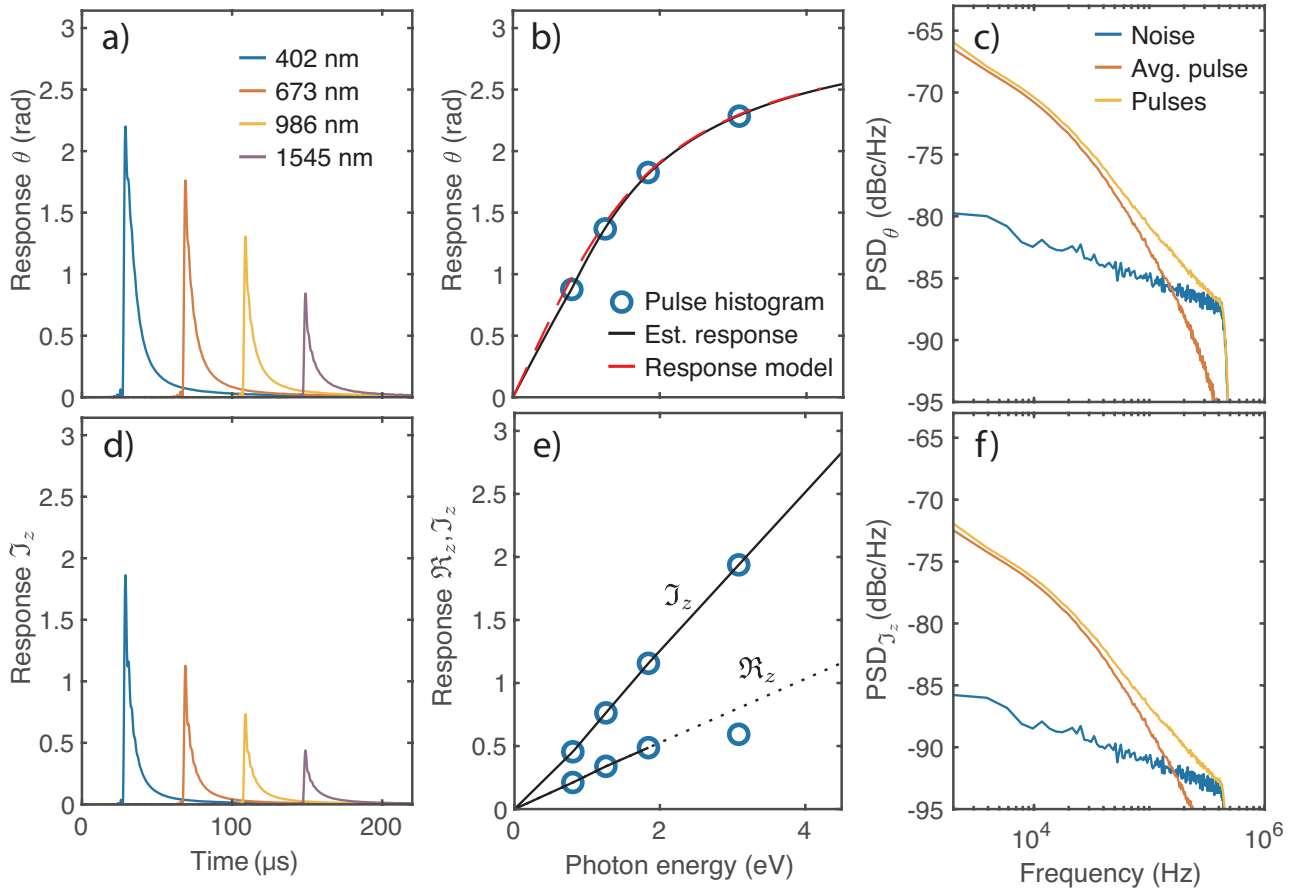}
  	\caption{Pulse information for both coordinate systems $(A, \theta)$ and ($\mathfrak{R}_z$, $\mathfrak{I}_z$). The corresponding photon energies are 3.08 eV for 402 nm, 1.84 eV for 673 nm, 1.26 eV for 986 nm, and 0.80 eV for 1545 nm. \textbf{a) and d)} Average pulse of $\sim$ 1000 pulses, used as the pulse shape for the optimal filter. \textbf{b) and e)} KID response against photon energy. The measured average pulse height, mean of the histogram, at each laser wavelength (dots) plotted against the known laser photon energy. For the phase coordinate in the top row: in red, the modelled phase responsivity as discussed in Appendix \ref{supp:responsivity} fitted for one free parameter, the pair breaking efficiency $\eta_{pb}$. In black the $\theta$ responsivity retrieved from the linear ($\mathfrak{I}_z$,$\mathfrak{R}_z$) coordinates. The modelled phase responsivity includes the free fit parameter $\eta_{pb}$. For the alternative coordinates in the bottom row: linear interpolated responsivity for $\mathfrak{I}_z$ and $\mathfrak{R}_z$. We see saturation of $\mathfrak{R}_z$ for the highest pulses in the 402 nm measurement, see section~\ref{subsec:comb_est}. \textbf{c) and f)} Typical power spectral densities of the average noise (blue), spectrum of the averaged pulse shape (orange), and the average spectrum of the pulses (yellow) for a 1545 nm laser measurement. }
  	\label{fig:pulse_prop}
\end{figure*}

We record an IQ timestream of 40 seconds at 1 Msample/s with enough laser attenuation to get roughly 40 photons per second.
Each pulse is cut from the timestream with a window width of 512 points, corresponding to roughly 0.5 ms.
Additionally, we record 20 seconds of data with the laser switched off to estimate the noise without the presence of photon hits $N(f)$.
Although we assume the pulse shape in Eq.~\ref{eq:op_filter} to be independent of the photon energy, this is not always the case.
Therefore, to characterize the detector at specific wavelengths, the pulse shape $M(f)$ is constructed by taking the average of a few thousand photon hits. 
The average pulses for each wavelength, used as the pulse shape at that wavelength, are plotted in Fig.~\ref{fig:pulse_prop}a.

Care has to be taken to properly align the pulses on arrival time.
We align the pulses on the rising edge, which we define as the point where the pulse first crosses half the pulse height.
However, with a rise time in the order of 1 $\mu$s and a sample rate of 1 Msample/s the rising edge is sampled at just 1 or 2 points.
To improve the pulse alignment we upsample the pulses, typically with a factor 8, improving the photon arrival time estimation to better than the sampling rate.
An overview of the upsampling procedure is given in Appendix~\ref{supp:upsample}.  

\begin{figure}[h!]
  	\includegraphics[width = \linewidth]{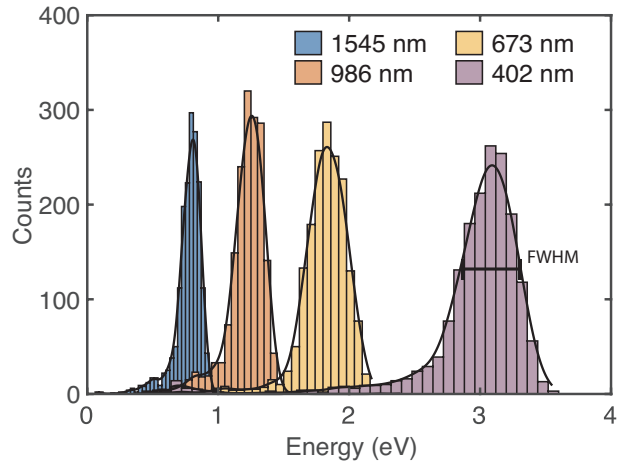}
  	\caption{Histograms measured individually for all four laser wavelengths together with the kernel density estimates used to determine the resolving power. The corresponding photon energies are 3.08 eV for 402 nm, 1.84 eV for 673 nm, 1.26 eV for 986 nm, and 0.80 eV for 1545 nm. The number of pulses in each histogram are limited to 1718 points, to have the same number of pulses in each histogram. The resolving power at each wavelength is given in Table~\ref{tab:R}.}
  	\label{fig:histograms}
\end{figure}

We apply the optimal filter to each of the individual pulses, yielding a few thousand estimates of the pulse height $\hat H(E_{ph})$.
The pulse height estimates are translated to photon energy estimates by the KID responsivity, see Fig.~\ref{fig:pulse_prop}b.
This yields the four histograms, for four laser wavelengths, in Fig.~\ref{fig:histograms}.
For some of the KIDs, we see a distinct low energy tail in the histogram as in Fig.~Fig.~\ref{fig:histograms}, which is absent for other KIDs, without a clear correlation with design, front/back illumination, or lens coupling.
An example of a histogram without a low energy tail is given in Appendix~\ref{supp:tail}, Fig.~\ref{fig:e_tail}.

We estimate the normal probability density function of each histogram with a kernel density estimation, which yields the resolving power of the detector as $R = \mu/$FWHM with $\mu$ is the mean and FWHM the full width half maximum of the distribution.
The estimated resolving powers depend on the chosen kernel width.
Varying the kernel width such that the density estimate gives either a too coarse or too fine representation of the histogram shows that there is a $\pm$5$\%$ uncertainty margin in the estimated resolving power.

All KIDs, whether lens-coupled or not, show a $R$ that saturates at $\sim$5-6 for 1545 - 402 nm.
The resolving power saturates at the point where the phase response saturates, which in most cases already starts at 1545 nm.
A detailed discussion of the saturated phase response is presented in Section~\ref{sec:resp} while a detailed analysis of the limits in $R$ is presented in Section~\ref{subsec:Res_Pow}.
\color{black}

\subsubsection{Responsivity}
\label{sec:resp}

We estimate the responsivity of the KID by recording the mean of $\hat H$ for the four known laser wavelengths at 402, 673, 986, and 1545 nm.
For devices with a phase responsivity such that pulses do not exceed $\theta$ = $\pi/2$ radians, the response curve will be approximately linear.
However, for the $\beta$-Ta devices described here, pulses easily exceed $\pi/2$ for the shorter wavelengths resulting in saturation in phase response \cite{zobrist_improving_2021}, see Fig.~\ref{fig:pulse_prop}b.

The saturation is a direct consequence of the polar coordinate system used to map the KID response into the conventional phase and amplitude response.
This effect can be reproduced by a response model based on the number of excess quasiparticles $N_{qp}$ generated by a photon, see Fig.~\ref{fig:pulse_prop}.
A detailed description is presented in Appendix~\ref{supp:responsivity}.

\color{black}
An alternative coordinate system presented in \cite{zobrist_improving_2021} provides a response that is linear in photon energy.
This coordinate system is analogous to the definition of the Smith chart, see Appendix \ref{supp:coord}, and is linear and monotonic in both coordinates,
In this work we adopt the notation $\mathfrak{I}_z$ for the new phase component, and $\mathfrak{R}_z$ as the new dissipation component.
A comparison between the two coordinate systems is presented in Fig.~\ref{fig:pulse_prop}.

\subsubsection{Resolving power contributions}
\label{subsec:Res_Pow}

First, we investigate the resolving power in the phase coordinate, see Table~\ref{tab:R}.
The phase response saturates for photon energies $\gtrapprox$ 1.28 eV, which limits the obtained resolving powers by reducing the signal-to-noise ratio and, above all, introduces an error in the optimal filter.
The error is twofold: saturation causes an energy-dependent pulse shape $M(f, E_{ph})$ and compresses the observed TLS-noise fluctuations, which scale with the resonator's response, causing an energy-dependent noise contribution $N(f, E_{ph})$.
Since both the pulse shape and the noise contribution are now energy-dependent, the modeled pulse $\hat{D}(f) = H(E_{ph})M(f)+N(f)$ used in the optimal filter will no longer be a valid representation of the measured pulse $D(f)$, see eq.~\ref{eq:op_filter}.

To estimate the effect of the response saturation on the resolving power, we analyse the measured resolving power at 1545 nm where the response is still linear, and compare it to the resolving power measured at shorter wavelengths.
For this analysis we select a KID for which the full 1545 nm histogram falls in the linear phase response regime.
The KID that fits this description is a front-illuminated KID without lens of which the measured resolving power is presented in Table~\ref{tab:R}.

\begin{table}[h!]
\caption{Resolving power ($E/\delta E$) for the different coordinate systems $(A,\theta)$ and $(\mathfrak{R}_z,\mathfrak{I}_z)$ obtained from the histogram mean and FWHM, see Fig.~\ref{fig:histograms}. The manual selection of the kernel width for the kernel density estimate gives the uncertainty margin on the estimated resolving powers, see Section~\ref{sec:Meas}. $R(\mathfrak{R}_z,\mathfrak{I}_z)$ is the resolving power for the combined estimator of the photon peak in both coordinate systems, see section~\ref{subsec:comb_est} and Appendix~\ref{supp:combined}.}
\label{tab:R}
	\begin{ruledtabular}
		\begin{tabular}{lcccc}
			\textrm{$\lambda$}&
			\textrm{$R(\theta)$}&
			\textrm{$R(\mathfrak{I}_z)$}&
			\textrm{$R(\mathfrak{R}_z)$}&
			\textrm{$R(\mathfrak{R}_z, \mathfrak{I}_z)$}\\
			
			\colrule
			1545 nm 	& 4.6 $\pm$ 0.23 & 4.7 $\pm$ 0.24 & 4.9 $\pm$ 0.25 & 5.0 $\pm$ 0.25	\\
			986 nm 	& 4.7 $\pm$ 0.24 & 4.8 $\pm$ 0.24 & 5.0 $\pm$ 0.25 & 5.2 $\pm$ 0.26	\\
			673 nm	& 4.6 $\pm$ 0.23 & 5.0 $\pm$ 0.25 & 5.5 $\pm$ 0.28 & 5.6 $\pm$ 0.28	\\
			402 nm 	& 4.9 $\pm$ 0.25 & 5.9 $\pm$ 0.30 & -\footnote{The dynamic range in these coordinates is limited, see main text.}	& -$^{\text{a}}$	\\
		\end{tabular}
	\end{ruledtabular}
\end{table}

For the $(A,\theta)$ coordinate, we do not present data in amplitude since the response is not only saturated but non-monotonic, which introduces an uncertainty in the translation from pulse height to photon energy.

At 1545 nm, in the phase coordinate, we measure a $R=$ 4.6 $\pm$ 0.23 and a $R_{SN}$ of 13.4,

\begin{equation}
	R_{SN} = \frac{\bar{H}}{2\sqrt{2\ln2}}\sqrt{\int_{-\infty}^{\infty}\frac{|M(f)|^2}{|N(f)|^2}df}.
	\label{eq:Rsn}	
\end{equation}
 
We can define $R_i$ as a measure of the intrinsic resolving power contributions that are not due to noise \cite{de_visser_phonon-trapping-enhanced_2021} using $1/R_{i}^2 = 1/R^2 - 1/R_{SN}^2$.
This gives an $R_i$ of 4.9$\pm$0.28 at 1545 nm which indicates that this KID is limited by intrinsic effects, probably hot phonon loss where phonons escape to the substrate without creating quasiparticle excitations.

The definition of $R_i$ is only valid for a linear response, which is only true for the measurement at 1545 nm.
We can however extend this analysis to the shorter wavelengths under the assumption that phonon losses limit the energy resolution, using the know relation

\begin{equation}
	R_{phonon} = \frac{1}{2\sqrt{2\ln2}}\sqrt{\frac{\eta_{pb}E_{ph}}{\Delta(F+J)}},
	\label{eq:Rphonon}
\end{equation}

where the phonon-loss is represented by the additional factor $J$.

The phonon loss factor $J$ depends on the depth at which photons are absorbed, since photons absorbed near the metal-substrate interface create phonons with a high probability to escape to the substrate.
Since the electromagnetic penetration depth ranges from 45 nm at 1545 nm to 28 nm at 402 nm for the 60 nm $\beta$-Ta film, $J$ can be energy dependent.
If this is a dominant effect, a back- instead of frontside illuminated device will have a lower and energy dependent $J$.
To verify that the analysis presented here is representative for the backside illuminated devices, we compare several KIDs illuminated from the front- and backside.
We do not see a difference in the obtained $R_i$.

Since the measured resolving power, $R(\theta)$ in Table~\ref{tab:R}, does not follow a $\sqrt{E_{ph}}$ dependence and we do not see a difference in back- or front-side illumination, there must be an additional mechanism that limits the resolving power at higher energies.
We attribute this mechanism to the saturation in phase response.

Based on the measured values of $R=$ 4.6 $\pm$ 0.23, $R_{SN}=$ 13.4 and $R_i=$ 4.9 $\pm$ 0.28; and the expected $E_{ph}$ dependancies of $R_{SN}$ and $R_i$ we estimate a phonon-limited $R$ of 9.6$\pm$0.54 at 402 nm.
The saturated response then limits the resolving power to the measured 4.9$\pm$0.28.

The linear Smith chart coordinates should then remove the error from the energy-dependent pulse shape $M(f)$ and noise contribution $N(f)$ caused by the saturated response.
We do indeed see an increase in the measured resolving power, from 4.9 to 5.9 at 402 nm.
However, the resolving power shows a much weaker energy dependence than the expected phonon-loss dominated $\sqrt{E_{ph}}$ dependence, see Table~\ref{tab:R}.$R(\mathfrak{I}_z)$.
We attribute the deviation to the non-stationarity of the amplifier noise in the coordinates $(\mathfrak{R}_z, \mathfrak{I}_z)$, which is mapped differently at different response heights making the amplifier noise non-stationary during a photon pulse.
A way to alleviate the error introduced by the non-stationarity is to measure with a parametric amplifier as is done in \cite{zobrist_improving_2021}.

\subsubsection{Combined estimator}
\label{subsec:comb_est}
For the linear coordinate system $(\mathfrak{R}_z, \mathfrak{I}_z)$ it can be beneficial to use a combined estimator, which combines the photon pulse data from both coordinates to estimate the photon energy, see Appendix~\ref{supp:combined}.
The combined estimator raises $R$ slightly to 5.0 - 5.6 for 1545 - 673 nm, see Table~\ref{tab:R}.$R(\mathfrak{R}_z, \mathfrak{I}_z)$.
However, the $\mathfrak{R}_z$ coordinate does not give an accurate pulse shape for the highest pules in the 402 nm dataset. 
The pulses are flattened at the top and no longer match with the estimated pulse model.
This behaviour is not visible in the response model of Appendix~
\ref{supp:responsivity} combined with the equations from Appendix~\ref{supp:coord}.
This makes the resolving power measurement at 402 nm unreliable and we have therefore chosen to omit the values from Table~\ref{tab:R}.
So, even though the combined estimator gives the best resolving power, the dynamic range of this estimator is limited.

\subsubsection{Resolving two spectral lines}

\begin{figure}[h!]
	\includegraphics[width = \linewidth]{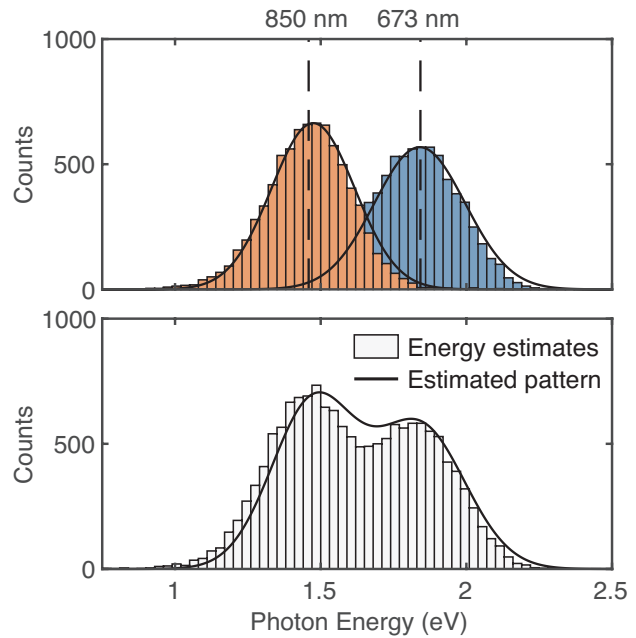}
  	\caption{\textbf{Top:} Histograms for separate laser (673 nm) and monochromator (850 nm) measurements. In this case the pulse shape is determined individually for both measurements. The solid line is the normal distribution corresponding to the measured resolving power and mean pulse height. \textbf{Bottom:} The energy estimates when the pulse shape determined for the laser measurement is applied to both laser (673 nm) and monochromator (850 nm) measurements. The solid line is the estimated pattern, obtained by combining the normal distributions from the top panel. The energy estimates are obtained with the combined filter in $(\mathfrak{R}_z, \mathfrak{I}_z)$, presented in Appendix \ref{supp:combined}.}
  	\label{fig:laser_and_mono}
\end{figure}

The monochromator and one of the four lasers can be coupled into the cryostat simultaneously so that we can measure the KIDs response to two lines.
The laser is set to 673 nm and the monochromator to 850 nm, which should be resolvable based on the measured resolving powers from Table \ref{tab:R}.
We measure the sources sequentially, taking 10-second timestreams for both the laser the monochromator.
First we analyse the two timestreams separately, determining the pulse shapes $M(f)$, average pulse height, and resolving power at both wavelengths, see Fig.~\ref{fig:laser_and_mono}.
The histograms in Fig.~\ref{fig:laser_and_mono} are obtained with the combined estimator in $(\mathfrak{R}_z, \mathfrak{I}_z)$ discussed in Appendix \ref{supp:combined} for which the KID response is linear in this wavelength range.

Next we process the entire 20-second timestream with the pulse shapes $M(f)$ obtained from the laser measurement. 
If the pulse shape is energy independent and the response is linear this should yield a histogram that is simply the sum of the laser and monochromator measurements. 
We can estimate the shape of this combined histogram by summing the normal distributions corresponding to the measured resolving power and mean pulse height of both sources, see Fig.~\ref{fig:laser_and_mono}.
The actual histogram, when all pulses are processed with the laser pulse shape of 673 nm, is presented in the bottom panel of Fig.~\ref{fig:laser_and_mono}.
The two lines can be separated from each other, as predicted from the individual measurements.
We conclude that the resolving powers found for single line measurements, as in Fig.~\ref{fig:histograms}, accurately describe the energy resolving capabilities of a KID.

The shape of the total histogram deviates from the expected shape, especially in the region of the monochromator measurement (850 nm).
The deviation from the expected pattern can then be explained by an energy-dependent pulse shape, such that the pulse shape determined at 673 nm is not fully representative of the pulse shape at 850 nm.
Comparing the pulse shapes at 673 and 850 nm, we do indeed observe this energy dependency in the pulse tail.

\section{Discussion}
\label{sec:Discussion}
The resolving powers in Table~\ref{tab:R} are representative of all the devices discussed in this work and represent the best estimate of the resolving power of these devices.
A dataset which contains pulse-data from all the measured devices, is available in the reproduction package of this paper, see Sec.~\ref{sec:Conclusion}.

The resolving power in the phase coordinate ($\theta$) saturates to $\sim$5 due to the saturation in phase response.
The linear coordinate $\mathfrak{I}_z$ from \cite{zobrist_improving_2021} and Appendix~\ref{supp:coord} raises the resolving power to 4.7 - 5.9 at 1545 - 402 nm. 
The combined estimator of Appendix~\ref{supp:combined} with both linear coordinates $(\mathfrak{R}_z, \mathfrak{I}_z)$ raises the resolving power slightly for a limited dynamic range.

\subsection{Saturated response}

The non-linear, saturated response, in KID phase, as presented in Fig.~\ref{fig:pulse_prop}b, limits the resolving power of these devices as discussed in Section \ref{subsec:Res_Pow}.
In Sec.~\ref{sec:resp} we discussed an alternative coordinate system that is linear in photon energy at the cost of non-stationary amplifier noise.
The measurements in \cite{zobrist_improving_2021} are performed with a parametric amplifier, reducing the amplifier noise level.
A lower amplifier noise level reduces the error the non-stationary noise introduces in alternative coordinate system $(\mathfrak{R}_z, \mathfrak{I}_z)$.
For KIDs operated without a parametric amplifier, we think the best way forward is to reduce the KIDs responsivity enough to get a linear KID phase response over the 1545 nm - 402 nm range.

Reducing the KID response does not come without a price as it lowers the obtained signal to noise resolving power $R_{SN}$.
To reach a high resolving power, the reduction in KID response should therefore be compensated.
One option is to improve the power handling of the devices \cite{clem_geometry-dependent_2011} which will reduce the amplifier noise and TLS noise levels.
Alternatively, the pulse decay time might increase with phonon trapping \cite{de_visser_phonon-trapping-enhanced_2021}, raising the integrable energy (area) of a photon pulse thus improving the signal to noise ratio.

Even when the KID is operated in the linear response regime, phonon loss will still limit the resolving power as discussed in Section \ref{subsec:Res_Pow}. 
From the measurement at 1545 nm we estimate that the phonon loss will limit the resolving power to 9.6 at 402 nm. 
Phonon (re)trapping, either by placing the KID on a membrane \cite{de_visser_phonon-trapping-enhanced_2021} or on a phonic barrier \cite{zobrist_membraneless_2022}, can improve the phonon-loss intrinsic resolving power $R_i$.

\subsection{Lifetimes}

\begin{figure}[hbt]
	\includegraphics[width = \linewidth]{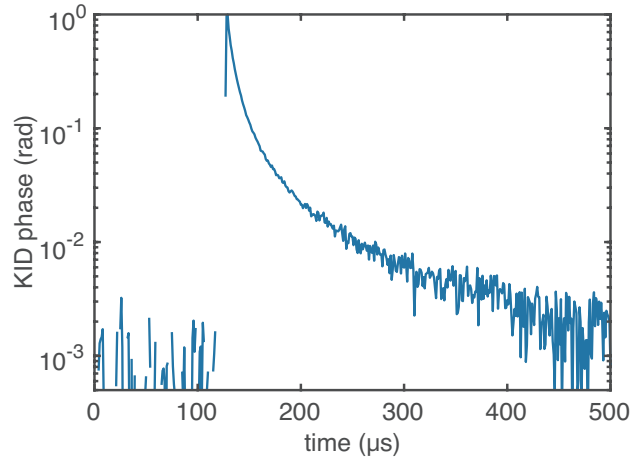}
  	\caption{Phase ($\theta$) pulse at 1545 nm in logarithmic scale to highlight the two pulse decay times. The pulse shows two timescales, a fast initial $\approx$ 7 $\mu$s decay and a slower $\approx$ 70 $\mu$s decay.}
  	\label{fig:double_t}
\end{figure}

The photon pulses show two lifetimes, one fast initial decay of $\approx$ 7 $\mu$s and a slower second decay of $\approx$ 70 $\mu$s, see Fig.~\ref{fig:double_t}.
The origin of these two lifetimes is currently under study.
The slower decay limits the maximum photon count rate. The fast initial decay limits the integratable area underneath the pulse which limits the obtainable $R_{SN}$.
We will investigate how the lifetimes depend on the material growth under different sputter conditions.

Increased phonon trapping increased the lifetime in Al KIDs \cite{de_visser_phonon-trapping-enhanced_2021}, and will increase $R_i$ improving the resolving power of these devices.
We will test the effect of phonon trapping by making KIDs on a membrane \cite{de_visser_phonon-trapping-enhanced_2021} and on a material with different phonon properties, creating a phononic barrier \cite{zobrist_membraneless_2022}.

\subsection{Fabrication}

The current fabrication process consists of 5 steps, as described in Sec.~\ref{sec:Des_and_Fab}, where the last 2 steps are a resist layer to protect the $\beta$-Ta and the Al for readout- and coupling bridges.
If we change the metal for the readout bridges from Al to $\beta$-Ta we can reduce the process to 3 steps with $\beta$-Ta as the final layer.
Since there is no process step after the $\beta$-Ta, we expect a $Q_i$ that is at least as high as measured here.
We achieve a 100\% fabrication yield for the 6$\times$6 array, which is a positive sign for the development of bigger arrays in the future.
 
The 3D-printed lenses, described in detail in Appendix \ref{supp:Lens_alignment}, are an interesting option for rapid prototyping of lens-coupled devices.
However, with the resin we currently use, there was one isolated row of 6 lenses which came loose after 3 cooldown cycles.
This is likely due to the different thermal expansion coefficients of the lenses and the substrate.
More work is needed to investigate whether other resins, or an extra adhesion layer between substrate and lenses could solve this problem.
For larger arrays the printing speed needs to be optimized since the 18 lenses presented here take roughly a day of printing.

\section{Conclusion}
\label{sec:Conclusion}

We measure the resolving power of $\beta$-Ta hybrid LEKIDs which is limited to $R\sim$5 for 1545-402 nm by saturation in the KID's phase response.
The saturated, non-lineair, phase response distorts the pulse height histograms lowering the obtained resolving power.
The $\beta$-Ta devices, with a $T_c$ of 1.0 K and resistivity if 239 $\mu\Omega$cm, show an average $Q_i$ of 4.3$\times10^5$ $\pm$ 1.3 $\times10^5$.
For a small array of 6$\times$6 pixels, we get a 100\% fabrication yield.
The high internal quality factor $Q_i$, ease of fabrication, and the possibility of a 3-layer process make $\beta$-Ta/NbTiN hybrid devices a promising option for larger, kilo-pixel, KID arrays.
On the backside of the sapphire substrate we have 3D printed a micro-lens array, aligned to markers in the NbTiN layer.
The lenses are printed with an alignment accuracy of of $\delta_x = +5.8 \pm 0.5$ $\mu$m and $\delta_y = +8.3 \pm 3.3$ $\mu$m and estimated FWHM spot size, fitted with a Gaussian, of 6.2 $\pm$ 0.7 $\mu$m.

All presented data, raw data, and analysis scripts are made available in a reproduction package uploaded to \textbf{Zenodo}: \url{https://doi.org/10.5281/zenodo.6719956}.

\section{Acknowledgements}
We acknowledge Nick de Keijzer and Robert Huiting for their work on the 100 mK sample stage.

This work is financially supported by the Netherlands Organisation for Scientific Research NWO (Projectruimte 680-91-127)

\appendix
\section{Lens analysis}
\label{supp:Lens_alignment}

\begin{figure}[h]
	\includegraphics[width = \linewidth]{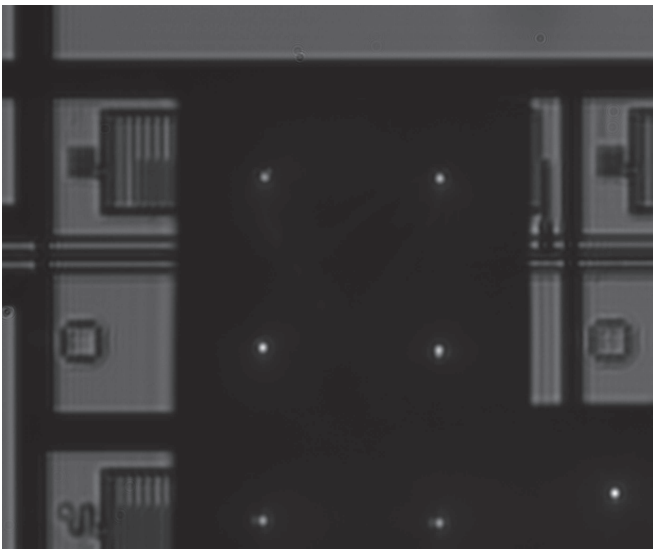}
	\caption{Re-imaged section of a LEKID test chip which contains a row of hybrid LEKIDs (top), alignment structures (middle), and LEKID design variations (bottom). Part of the CPW read-outline is visible between the top and middle rows. 6 lens focal spots are visible in the centre and a 7th spot, from a detached and shifted row, is visible in the lower right corner. The image is taken with a 3.5 $\mu$m pixel pitch CCD, with the array illuminated by a 673 nm laser.}
	\label{fig:lens_supp_A}
\end{figure}

We use a re-imaging setup \cite{haver_enabling_2010} to image the microlens focal spots on the KID layer.
The set-up consists of a monochromator source (673 nm), the microlens chip mounted on a manual single-axis z-stage, a re-imaging lens and a 3.45 $\mu$m pixel pitch CCD (Flir Blackfly BFLY-U3-50H5M-C).
The magnification of the setup is determined from the imaged CPW read-out line.
The lens misalignment is determined with respect to the imaged LEKID inductors and alignment structures as shown in Fig.~\ref{fig:lens_supp_A}.

The 3D printed micro lenses have an alignment error of $\delta_x = +5.8 \pm 0.5$ $\mu$m and $\delta_y = +8.3 \pm 3.3$ $\mu$m measured with respect to the desired focal spot location of 6 lenses.

In addition to the alignment error, we can measure the focal spot size by scanning the microlens array in the axial distance.
The spot size is determined from the depth profile of each focal spot.
The FWHM of the spot, fitted with a Gaussian, is 6.2 $\pm$ 0.7 $\mu$m.
The diffraction limited airy disk diameter at 673 nm for an aperture with $D$ = 150 $\mu$m and $F=525$ $\mu$m (the optical length in the dielectric) is 5.75 $\mu$m, and the FWHM is 2.42 $\mu$m.
The corresponding airy disk diameter for a FWHM of 6.2 $\mu$m is 14.2 $\mu$m.
For the measured alignment error and estimated Airy disk diameter the complete airy disk, which contains 83.8\% of the optical power, does not completely fall within the 23$\times$25 $\mu$m inductor.

\begin{figure*}[ht]
	\includegraphics[width = \textwidth]{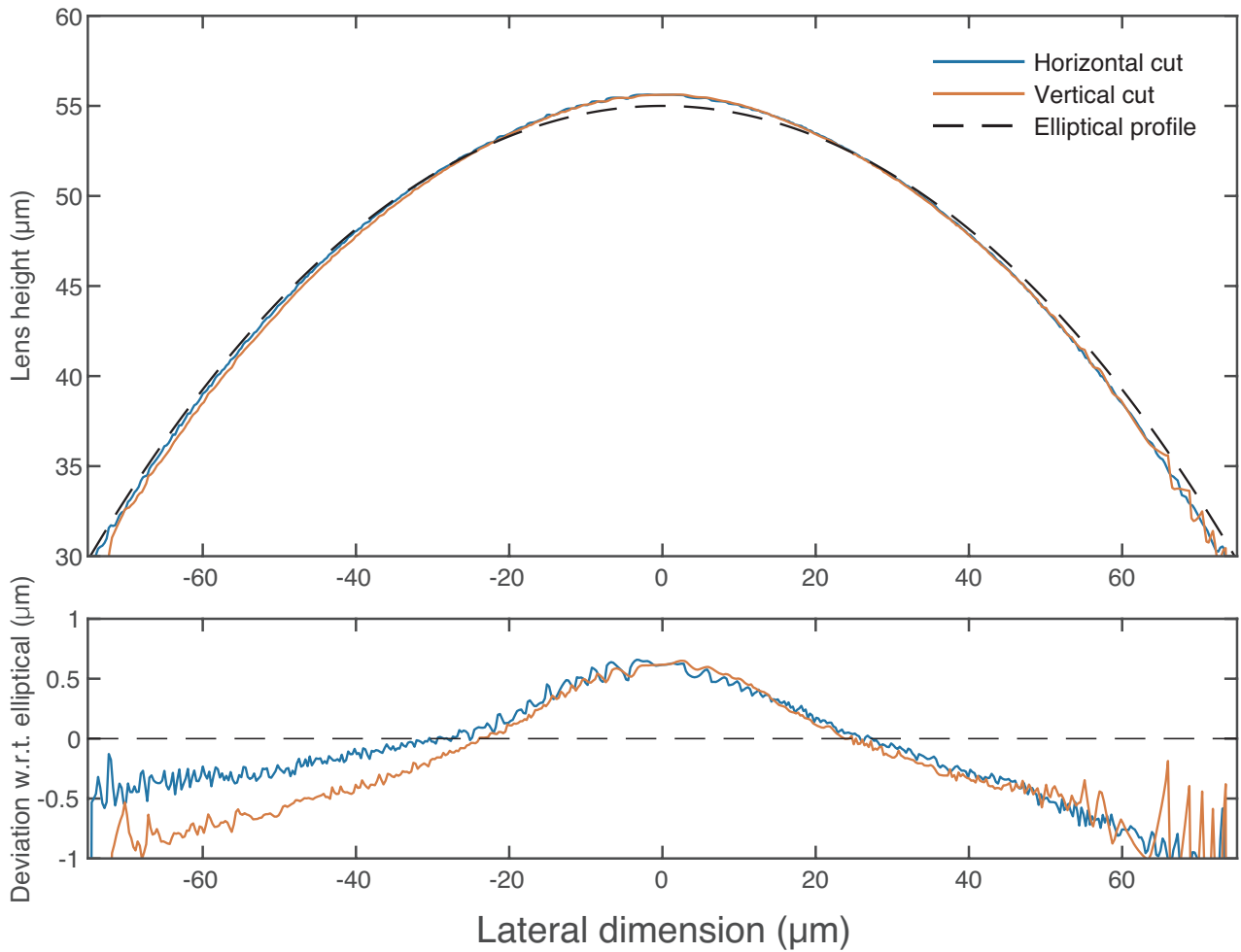}
	\caption{The measured horizontal and vertical cross-section of one of the Nanoscribe printed lenses together with the designed profile. The profile is measured with a Keyence VK-X1000 laser confocal microscope, using a laser-based height measurement.}
	\label{fig:lens_supp_B}
\end{figure*}

We characterise the shape of the lenses with a Keyence VK-X1000 laser confocal microscope, using a laser-based height measurement.
We measure height profiles of isolated lenses, printed with the same recipe.
The height profile measurement requires a height reference (a free view on the substrate surface) on at least 3 places around the lens within the field-of-view, which is not possible on a lens-array.
The measured horizontal and vertical cross-section of one of the lenses is shown in Fig.~\ref{fig:lens_supp_B}, together with the designed profile.
The bottom panel shows that the difference between design and realisation is better than 1 $\mu$m.

The final recipe for the Nanoscribe printer is the result of an optimization process where we have varied the scan speed and laser intensity of the printer, which together with the choice of microscope objective set the effective dose with which the resin is cured.
Each iteration was evaluated with a height-profile measurement to find the conditions where the designed profile was most accurately represented.
We start the writing process 1 $\mu$m below the substrate surface to ensure adhesion. 

\section{Upsampling}
\label{supp:upsample}
\begin{figure}[h!]
  \includegraphics[width=\linewidth]{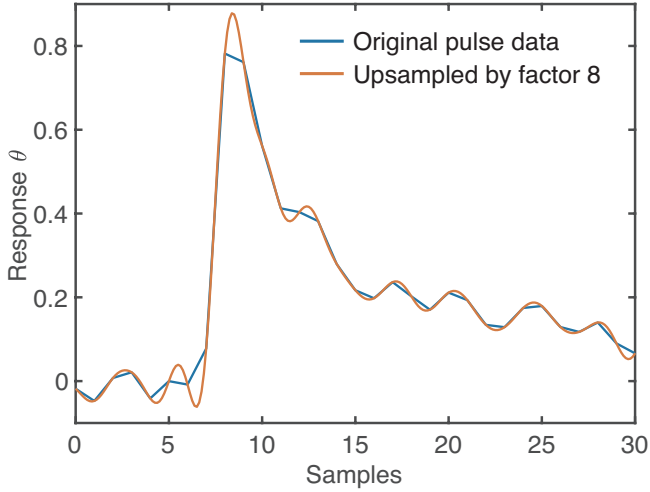}
  \caption{Original and factor 8 upsampled single photon pulse in KID phase response $\theta$. Upsampled with Matlab resample function, which uses a FIR antialiasing lowpass filter.}
  \label{fig:upsample}
\end{figure}

The pulses are aligned based on their rising slope, taken as the point where the pulse first crosses 0.5 times its final height.
However, with a sample-rate of 1 Msample/s and a rise time of $\tau_{res} = 2Q/\omega_0 \approx$ 1 $\mu$s, the rising edge is sampled at one or two points.
To better estimate the rising edge we upsample the pulse time-stream by an integer factor, see Fig.~\ref{fig:upsample}, improving the photon arrival time estimation to better than the sampling rate.
The upsampling operation is done with the resample Matlab function, which resamples the input sequence at x times the original sample rate.
The resample function applies a Finite Impulse Response (FIR) antialiasing lowpass filter to the input sequence and compensates for the delay introduced by the filter.
The original pulse window has a length of $2^n$ to improve the performance of Matlab's FFT function, so we use a upsample factor that is a power of two to make sure that the upsampled pulse has a $2^n$ length.

The resample function is sensitive to large transients in the input signal.
The photon pulses contain two of these transients, at the start of the peak where the signal rises sharply and at the under sampled peak where the signal drops suddenly.
At these points the filter of the resample function slightly overestimates the actual pulse shape, see Fig.~\ref{fig:upsample}.
This behaviour is however consistent for all the pulses and does not negatively affect the final energy resolution.
The improved photon arrival time estimations results in a better peak-to-peak alignment, which in turn improves the estimated pulse model and the pulse-to-model alignment for the optimal filter peak height estimation.

\section{Coordinate systems}
\label{supp:coord}

We use a homodyne readout scheme to measure the KIDs' microwave response to an excitation by a photon absorption event.
The output of the mixer is an in-phase component $I$ and a quadrature component $Q$ which are proportional the complex transmission coefficient $S_{21}$.
A measurement of a KID frequency sweep at constant microwave power traces a circle in the complex IQ plane.
After we calibrate for the cable delay and amplification in the readout chain, we translate the circle such that a frequency sweep measured in the dark, with the laser switched off, would trace a circle with radius 1, centered at (0,0). 
A noise measurement with the reference tone at the KIDs resonance frequency is then located at (-1,0).

When analyzing the response to radiation, $I$ and $Q$ are usually mapped on a polar coordinate system.
The phase response $\theta$ is the clock-wise angle with respect to the negative x-axis.
The dissipation response $A$ is the distance from the circle's centre at (0,0), we typically plot $\delta A = 1-A$.

The authors of \cite{zobrist_improving_2021} show that these coordinates are linear for small signals.
However, for larger signals the response becomes non-linear, and in the case of $A$ it becomes non-monotonic.
The authors then propose an alternate, Smith chart like, coordinate system $(\mathfrak{R}_z, \mathfrak{I}_z)$ that has a linear response to photon energy.

For a calibrated KID circle centered at (0,0) with a radius of 1, the equations for these new coordinates are given by

\begin{equation}
	\Gamma = I + iQ
	\label{eq:Gamma}
\end{equation}

\begin{equation}
	z = \frac{1+\Gamma}{1-\Gamma}
	\label{eq:z}
\end{equation}

The new coordinate system $(\mathfrak{R}_z, \mathfrak{I}_z)$ is given by the real and imaginary part of $z$ respectively.

For an asymmetric resonance dip, caused by a mismatch in the transmission line, the KID circle is rotated by $\phi$ and magnified by a factor $1/\cos(\phi)$ \cite{khalil_analysis_2012}.
The rotation and factor $1/\cos(\phi)$ have to be corrected before equations \ref{eq:Gamma} and \ref{eq:z} can be used.
Alternatively equations 10a and 10b in \cite{zobrist_improving_2021} provide a way to compensate for the asymmetry through the factor $x_a$. The factor $x_a$ relates to $\phi$ as

\begin{equation}
	x_a = \frac{Q_i}{Q_i+Q_c}\frac{\tan(\phi)}{2Q}.
\end{equation}
 
 Both methods give the same response except for a scaling factor.

 Until now we have assumed that one measures on resonance, where the read-tone is equal to the KID resonance frequency $f_0$.
 Measuring off-resonance, which could happen when $f_0$ is taken from a Lorentzian fit to an asymmetric resonance dip, effectively rotates the pulse trajectory along the KID circle.
 Since the Smith chart is rotationally stable, the off-resonance measurement just adds an offset to the Smith chart response which can easily be corrected for in post-processing.
 
  The reproduction package accompanying this paper contains several scripts that demonstrate the Smith chart like coordinate system and compare the obtained response with the formalism adopted in \cite{zobrist_improving_2021}.


\section{Combined estimator}
\label{supp:combined}

The pulse model discussed in the main text can be extended to include the photon pulse in both phase and dissipation coordinates.
The maximum likelihood estimator, in frequency domain is

\begin{equation}
	\chi^2 = \int_{-\infty}^{\infty}(D - HM)^* S^{-1}(D - HM) df,
\end{equation}

With $D$, $H$ and $M$ the column vectors containing the measured pulse, pulse height and pulse shape in both coordinates respectively, and $S$ the 2$\times$2 noise covariance matrix.
The best estimate of $E_{ph}$ is found by minimizing $\chi^{2}(E_{ph})$.

\section{Responsivity model}
\label{supp:responsivity}

In order to calculate the phase and amplitude response to an incoming photon, the equilibrium number of quasiparticles in the sensitive volume is calculated with, $N_{qp}^0 = 2 V N_0 \sqrt{2\pi k_B T_{bath} \Delta} e^{-\Delta/k_B T}$. 
Here, $N_0$ is the density of states at the Fermi energy, $V$ is the sensitive superconducting volume, $k_B$ is the Boltzmann constant, $\Delta$ is the superconducting order parameter and $T_{bath}$ is the bath temperature.
The equilibrium complex conductivity is calculated with the Mattis-Bardeen equations \cite{mattis_theory_1958} at $T_{bath}$.
With the complex conductivity, the internal quality factor, $Q_i$ is calculated. 
Together with the resonance frequency, $f_{res}$, and coupling quality factor, $Q_c$, the equilibrium $S_{21}$ circle in the complex plane is determined.

The number of excess quasiparticles generated by the photon is calculated via $\delta N_{qp} = \eta_{pb}\hbar\omega\Delta^{-1}$.
Here, $\hbar\omega$ is the photon energy and $\eta_{pb}$ is the pair breaking efficiency, which is the only fit parameter of the model and gives a reasonable value of 0.55.
Note that this fitted value of $eta_{pb}$ should not be interpreted as a measurement of a universal $eta_{pb}$. It shows that the saturation phenomena observed in the KID response can be consistently described with this simple model.

The number of quasiparticles just after the photon energy is absorbed, $N_{qp} = N_{qp}^0 + \delta N_{qp}$, is translated to an effective quasiparticle temperature, $T_{eff}$, by inverting the equation for $N_{qp}^0$.
From $T_{eff}$, the new complex conductivity is calculated with the Mattis-Bardeen equations \cite{mattis_theory_1958}. 
From this, the new $Q_i$ and $f_{res}$ of the KID are calculated.
With these variables, the complex scattering parameter $S_{21}$ is calculated, which is translated to an amplitude and phase, relative to the equilibrium $S_{21}$ circle.
For details, see \cite{de_visser_quasiparticle_2014, de_rooij_quasiparticle_2020}.

We limit ourselves here to a model of the pulse-height only.
The pulse-decay could be added to the model starting with the Rothwarf-Taylor equations \cite{rothwarf_measurement_1967}.
The comparison of such model to the observed quasiparticle dynamics is not straightforward, and beyond the scope of this work.

\section{Low energy tail}
\label{supp:tail}
\begin{figure}[hbt]
	\includegraphics[width=\linewidth]{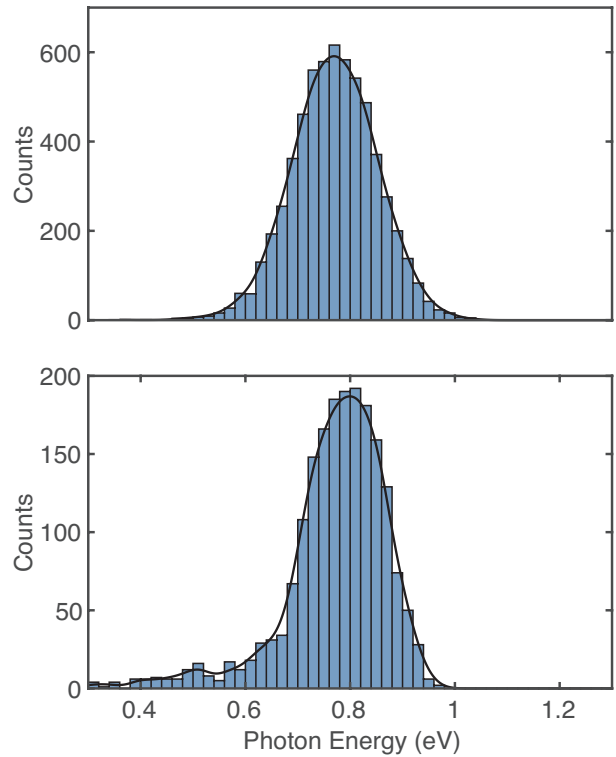}  
  	\caption{Histograms at 1545 nm for two different KIDs with the same design. The top histogram shows a symmetric distribution while the bottom histogram shows a significant low energy tail.}
  	\label{fig:e_tail}
\end{figure}

For each laser measurement the photon energy estimates are presented as histograms, see Fig.~\ref{fig:histograms}.
For some of the KIDs, we see a distinct low energy tail in the histogram, which is absent for other KIDs, without a clear correlation between design, front/back illumination, or lens coupling. Two examples of these histograms, one with and one without a low energy tail, are presented in Fig.~\ref{fig:e_tail}.

Such a low energy tail is common for non-hybrid KIDs and is typically explained by quasiparticle leakage into the less sensitive capacitor \cite{zobrist_membraneless_2022}.
This should not be an issue for hybrid KIDs due to the difference in energy gap of the low-$T_c$  $\beta$-Ta inductor and the high-$T_c$ NbTiN capacitor. 
Further study is required to pinpoint the source of this low energy tail in hybrid LEKIDs.

\bibliography{ref}

\end{document}